\documentclass[aps,twocolumn,pre]{revtex4}
\usepackage[dvips]{graphics}
\usepackage{graphicx}
\usepackage{amsfonts}
\usepackage{amssymb}
\usepackage{amsmath}

\def\wfig{4.25cm}
\def\wwfig{5.2cm}

\begin{document}

\title{Spontaneous Symmetry Breaking in Photonic Lattices: Theory and
Experiment}
\author{P.G. Kevrekidis$^1$, Zhigang Chen$^2$, B.A. Malomed$^3$, D.J.\
Frantzeskakis$^{4}$ and M.I. Weinstein$^5$ }
\affiliation{$^1$ Department of Mathematics and Statistics, University of Massachusetts,
Amherst MA 01003-4515, USA \\
$^{2}$ Department of Physics and Astronomy, San Francisco State University,
CA 94132, and TEDA College, Nankai University, Tianjin, 300457 China\\
$^{3}$ Department of Interdisciplinary Studies, Faculty of Engineering, Tel
Aviv University, Tel Aviv 69978, Israel \\
$^4$ Department of Physics, University of Athens, Panepistimiopolis,
Zografos, Athens 15784, Greece \\
$^5$ Department of Applied Physics and Applied Mathematics Columbia
University, New York, NY 10025, USA and Mathematical Sciences Research, Bell
Laboratories, Murray Hill, New Jersey, 07974, USA. }

\begin{abstract}
We examine an example of spontaneous symmetry breaking in a
double-well waveguide with a symmetric potential. The
ground state of the system beyond a critical power becomes
asymmetric. The effect is illustrated numerically, and
quantitatively analyzed via a Galerkin truncation that clearly
shows the bifurcation from a symmetric to an asymmetric steady
state. This phenomenon is also demonstrated experimentally when a
probe beam is launched appropriately into an optically induced
photonic lattice in a photorefractive material. 
\end{abstract}

\maketitle

%\textwidth 15.5cm
%\email{kevrekid@math.umass.edu}

%\section{Introduction\label{sec1}}

\textit{Introduction}. Spontaneous symmetry breaking (SSB) is a
ubiquitous phenomenon in modern physics. Manifestations of the SSB
have been found in diverse areas, ranging from liquid crystals
\cite{liquid} to quantum dots \cite{dots}, and from coupled
semiconductor lasers \cite{lasers} to the pattern dynamics of
\textit{Dictyostelium discoideum} \cite{turing}. Of particular
interest is the recent experimental demonstration of spatial
symmetry-breaking instability in the interaction of laser beams in
optical Kerr media \cite{HaeltermannPRL02}. For an overview of the
time-honored history of the SSB in field theory, see
the reviews \cite{archive}.

There has recently been a huge amount of activity and many
advances in the study of light dynamics in photonic
structures, such as materially fabricated photonic crystals (PCs)
and optically induced photonic lattices in nonlinear media; see,
for example, \cite{joannopoulos,kivshar}. This is motivated by the
enormous potential for applications ranging from highly tunable
telecommunications elements to cavity QED experiments.
Among many phenomena explored are nonlinear effects
associated with propagation, localization, and discretization of
light in optically-induced photonic lattices \cite{NatureDNC03},
including the formation of lattice solitons in one
\cite{FlescherPRL,NeshevOL} and two dimensions
\cite{FleischerNature03,MartinPRL04,ChenPRL04}, and discrete vortex
solitons \cite{NeshevPRL04,FleischerPRL04}.

In this Letter, our aim is to study a prototypical example of SSB
for a setting of the nonlinear Schr\"{o}dinger (NLS) type
with an effective symmetric double-well potential,
i.e., a prototypical ``dual-core'' photonic lattice. As the
optical power is increased, we identify a
symmetry-breaking bifurcation in the system's ground state, with
a transfer of stability to asymmetric states, with more power
concentrated in one core than in the other. This,
generally, resembles theoretically predicted SSB bifurcations in
diverse dual-core optical systems of more traditional types, such
as dual-core fibers (see Refs. \cite{Skinner}), linearly coupled 
$\chi ^{(2)}$ waveguides
and dual-core fiber gratings \cite{MakBragg}. We
present an analysis using a Galerkin truncation based on the
eigenfunction basis of the underlying linear double-well problem,
which accurately predicts the bifurcation in the present
context. Then, we demonstrate such SSB experimentally in
an optically-induced waveguide lattice in photorefractive media.
Our method of  prediction and analysis of SSB 
%applies
%in a general context. 
is quite general. For example, NLS equations of the
Gross-Pitaevskii and nonlinear-Hartree types play a
fundamental role in the study of Bose-Einstein
condensation \cite{BEC1}. In the latter context, a rigorous variational proof
showing that a SSB transition must occur is given in
\cite{frohlich}, and a complete study of the SSB transition in a
special one-dimensional (1D) model was developed in Ref.
\cite{jackson}.

The presentation is structured as follows. In section II, we introduce the
NLS model and its connection to the optical lattice problem. 
The stability of stationary states and the
SSB bifurcation are 
studied numerically. In section III, the finite-mode (Galerkin)
approximation and the prediction of SSB following from it are
elaborated. Section IV details the observation of the SSB in the
experiment. Section V contains a summary of our findings and conclusions.

\textit{Model and Numerical Results}. Our model is based on (1D)
equations that describe the propagation of light in a
photorefractive crystal \cite{EfremidisPRE02}; see also the recent
exposition in \cite{yang}. The 1D version makes it
possible to demonstrate the SSB in its simplest/most fundamental
form. Specifically, we consider a probe beam that is
extraordinarily polarized, while a strong ordinarily polarized
beam creates an effective lattice potential for the
probe. Then, the equation for the spatial evolution of a
slowly varying amplitude $U$ of the probe beam is
\begin{equation}
iU_{z}+\frac{1}{2k_{0}n_{e}}U_{xx}-\frac{1}{2}k_{0}n_{e}^{2}r_{33}\frac{E_{0}}{1+I_{0}(x)+|U|^{2}}U=0.
\label{pceq1}
\end{equation}In Eq. (\ref{pceq1}), $z$ and $x$ are the propagation distance and
transverse coordinate, respectively, $k_{0}$ is the wavenumber of the probe
beam in the vacuum, $n_{e}$ is the refractive index along the extraordinary
axis, $r_{33}$ is the electro-optic coefficient for the extraordinary
polarization, $E_{0}$ is the bias electric field, and $I_{0}(x)$ is the
intensity of the ordinarily polarized beam, subject to modulation in the
transverse direction (all intensities are normalized with respect to the
crystal's dark irradiance, $I_{d}$). Measuring $z$ in units of $2k_{0}n_{e}$
and $E_{0}$ in units of $1/(k_{0}^{2}n_{e}^{4}r_{33})$, Eq. (\ref{pceq1})
can be cast in a dimensionless form,
\begin{equation}
iU_{z}+U_{xx}-\frac{E_{0}}{1+I_{0}(x)+|U|^{2}}U=0.  \label{U}
\end{equation}
Nonlinear bound states of Eq. (\ref{U}) are localized solutions of
the form $U(x,z)=u(x)\exp (i\mu z)$, where $u$ obeys
\begin{equation}
u_{xx}-\frac{E_{0}}{1+I_{0}(x)+|u|^{2}}u=\mu u.  \label{pceq5}
\end{equation}

We consider the case of an effective symmetric two-hump potential,
\begin{equation}
I_{0}(x)=V_{0}\left[ \exp \left( -\frac{(x-a)^{2}}{2\epsilon ^{2}}\right)
+\exp \left( -\frac{(x+a)^{2}}{2\epsilon ^{2}}\right) \right] ,
\label{twohumps}
\end{equation}
corresponding to a superposition of two Gaussian beams. Solutions to Eq.
(\ref{pceq5}) with $I_{0}(x)$ given by (\ref{twohumps}) were found
via the Newton's method on a finite-difference grid. The linear
stability of the stationary states 
is determined by the eigenvalues and eigenvectors,
$\{\lambda ,\left( a,b\right) \}$ of the linearized equation,
obtained by the substitution of $U(x,z)=\exp (i\mu z)\left\{
u(x)+\delta \left[ \exp (-\lambda z)a(x)+\exp (-\lambda ^{\star
}z)b^{\star }(x)\right] \right\} $ into (\ref{pceq1}) and
linearization in the small parameter $\delta$.
 For very weak nonlinearity, the profile of the
ground state follows the symmetry of the double-well
potential (\ref{twohumps}). We searched for\ a symmetry-breaking
bifurcation  as the optical power 
of the nonlinear bound state, $N=\int_{-\infty
}^{+\infty }|u(x)|^{2}dx$ was increased.

\begin{figure}[t]
\centerline{
\includegraphics[width=\wfig,height=\wfig,angle=0,clip]{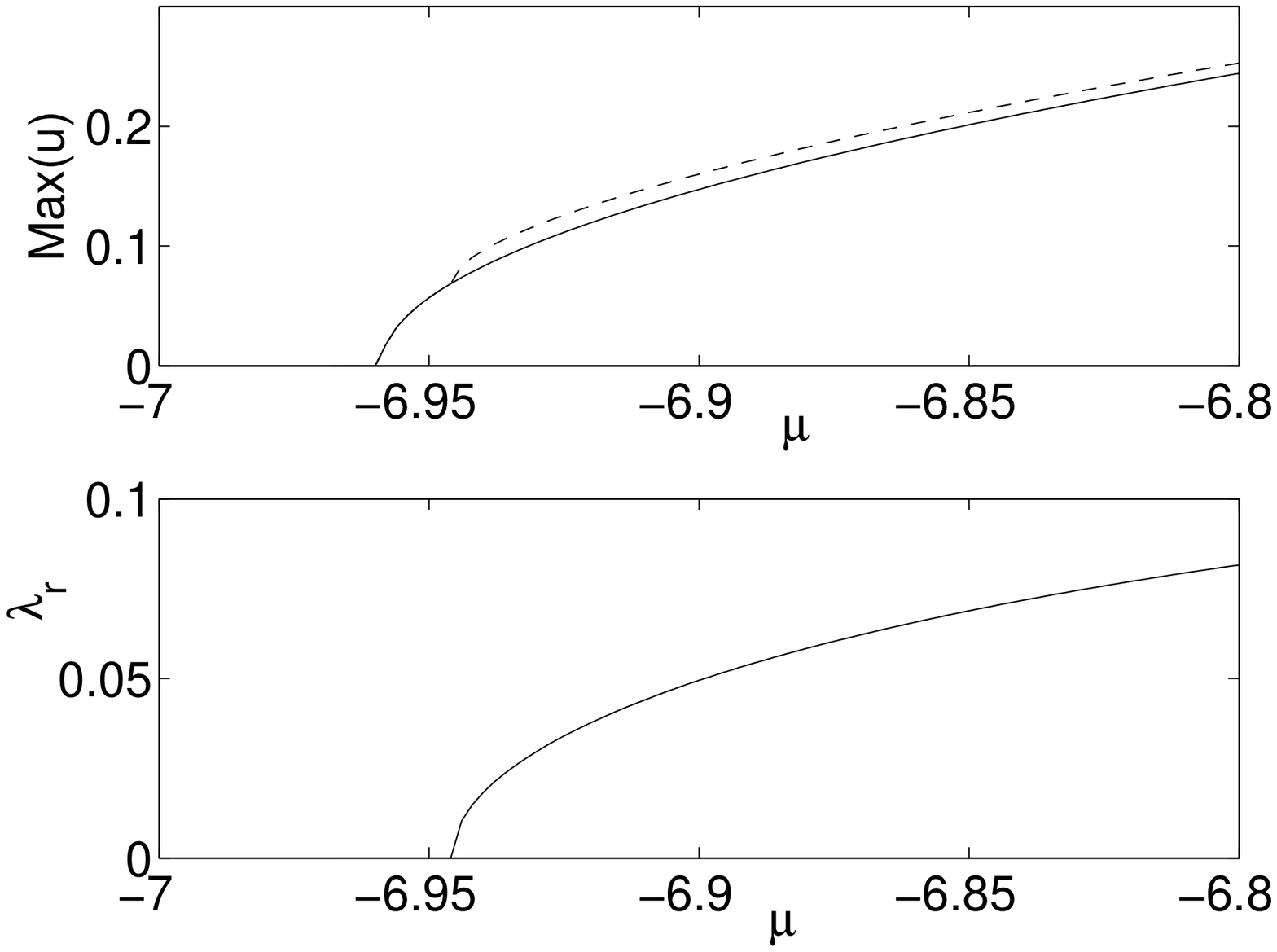}
\includegraphics[width=\wfig,height=\wfig,angle=0,clip]{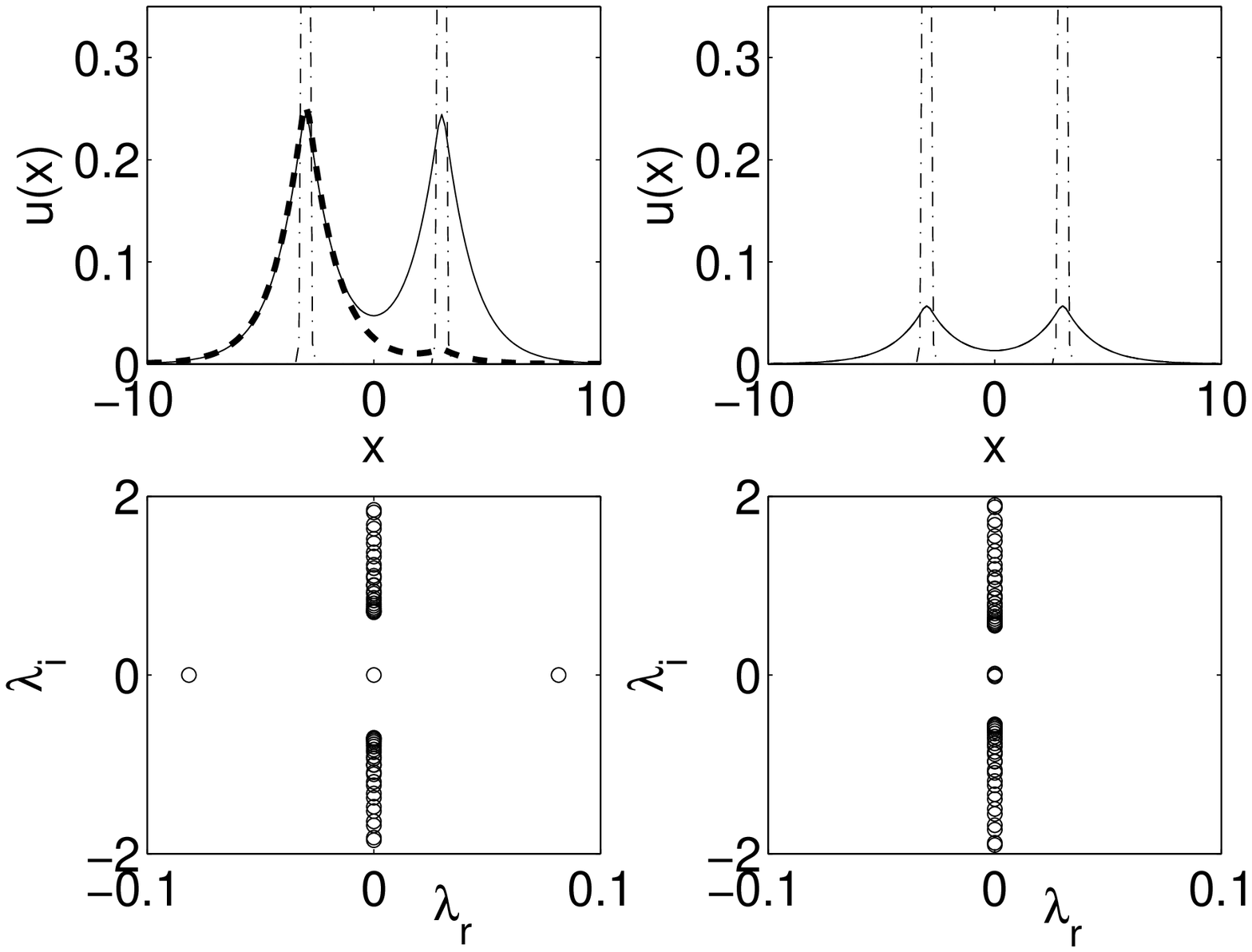}}
\caption{The symmetry-breaking bifurcation which gives rise to
asymmetric solutions in Eq. (\protect\ref{pceq5}) with
the potential of Eq. (\protect\ref{twohumps}).
The parameters are $\protect\epsilon =0.1$, $V_{0}=(\protect\pi
\protect\epsilon )^{-1/2}\approx 1.784$, $a=3$, and $E_{0}=7.5$.
The solid and dashed lines in the top subplot of the left panel
show the amplitudes of the symmetric and asymmetric (bifurcating)
solutions vs. $\protect\mu $. The bottom subplot displays the
unstable real eigenvalue for the symmetric-solution branch
existing past the bifurcation point, at $\protect\mu >-6.94575$.
The right panel shows, in its top subplots, examples of the
symmetric and asymmetric solutions (solid and thick dashed lines,
respectively) for $\protect\mu =-6.8$ (left) and $\protect\mu
=-6.95$ (right), respectively. In the latter case (i.e., before
the bifurcation point), only the symmetric branch (solid line)
exists. The background profile of $I_{0}(x)$ is indicated by a
dash-dotted line. The bottom subplots show the respective results of the
linear stability analysis around the symmetric solution in the
complex plane $(\protect\lambda _{r},\protect\lambda _{i})$ of the
stability eigenvalues. An eigenvalue with a positive real part (in
the left panel) implies instability of the solution.}
\label{pcfig1}
\end{figure}

The results are summarized in Fig. \ref{pcfig1} (for $E_{0}=7.5$, cf. Ref.
\cite{yang}). The top left panel of the 
figure shows the amplitude of the solution as a function
of $\mu $. The solid and dashed lines correspond, respectively, to the
symmetric and asymmetric solutions, the latter bifurcating from the former
(SSB) at a critical power $N_{c}$. 
%The top
%subplot of the right panel shows the amplitude of the solution, indicating
%the onset of an asymmetric state which concentrates more in one well than in
%the other. 
The bottom left panel displays the real part $\lambda _{r}$ of the most
unstable eigenvalue of the symmetric solution, which shows that the
symmetry-breaking bifurcation destabilizes the symmetric solution. This
%destabilization of the symmetric branch 
occurs through the appearance of a
pair of real eigenvalues for $N>N_c \equiv 0.15167$, or, 
equivalently, for $\mu >\mu_{c}\equiv -6.94575$, when the solution's 
amplitude (maximum value
of $\left\vert u(x)\right\vert $) exceeds $0.06924$. The asymmetric
solution emerging at the bifurcation point
is stable. The right panel of Fig. \ref{pcfig1} shows details of
the relevant solutions and their stability for $\mu > \mu_c$ (left)
and $\mu < \mu_c$ (right).
In fact, two asymmetric solutions arise, being
mirror images of each other, i.e., the bifurcation is super-critical.
%belongs to the well-known
%super-critical type.

\begin{figure}[t]
\centerline{
\includegraphics[width=\wwfig,height=\wwfig,angle=0,clip]{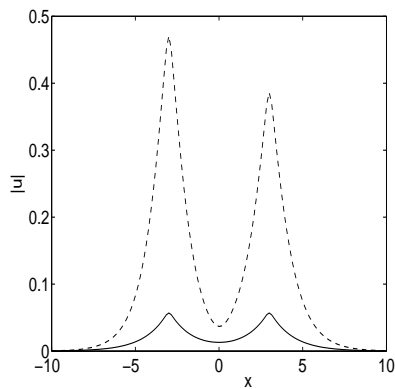}}
\caption{The profile of the field modulus $|u|$
for the same propagation distance but different symmetric initial
conditions: a stable one (solid line) for $\protect\mu =-6.95$, resulting in
the stable symmetric state, and an unstable one for
$\protect\mu =-6.5$, resulting in the asymmetric state
shown by the dashed line. The potential is the same as 
for Fig. \ref{pcfig1}.} 
\label{pcfig2}
\end{figure}

The evolution of the stable and unstable stationary
solutions was investigated in direct simulations of Eq.
(\ref{U}). Figure \ref{pcfig2} displays the results for the
unstable symmetric solution with $\mu =-6.5$ and stable symmetric
one with $\mu =-6.95$. As initial conditions, we took highly
accurate numerically obtained stationary states perturbed by
a uniformly distributed random perturbation of an amplitude
$0.0001$. In the unstable case, the manifestation of the
SSB is very clear; as a result of the growth of the unstable
eigenmode triggered by the small perturbation, the nearly
symmetric state with $\mu =-6.5$ evolves into a stable asymmetric
one; see the dashed line in Fig. \ref{pcfig2}. On the
other hand, the stable symmetric ground state for $\mu =-6.95$
remains unchanged (solid line in Fig. \ref{pcfig2}).

\textit{Analytical Results}. 
%\textbf{To develop an approximate analytical
To analytically examine SSB in
Eq. (\ref{U}), we develop a  
Galerkin-type method generally applicable to
related models. It is convenient to define $U(x,z)\equiv
u(x,z)\exp (i\mu z)$, replacing Eq. (\ref{U}) with
\begin{equation}
iu_{z}=\mu u-u_{xx}+\frac{E_{0}}{1+I_{0}(x)+|u|^{2}}u.  \label{pceq6}
\end{equation}
Numerics corroborate our expectation that,
for sufficiently large separation $a$ between the  potential wells,
the
bifurcation occurs at low powers, 
i.e.,
when Eq. (\ref{pceq6}) is close to its linear counterpart. 
It is
therefore natural to seek a representation of $u(x,z)$ in terms of
the the basis of the eigenfunctions of the double-well potential:
%\begin{equation}
$\omega_j v^{(j)}=-v_{xx}^{(j)}+\frac{E_{0}}{1+I_{0}(x)}v^{(j)}$.  
%\label{pceq7}
%\end{equation}
%\textbf{
Numerical solution of this equation 
%(\ref{pceq7}) \textbf{
reveals two
localized eigenstates of the double-well potential with 
corresponding eigenvalues $\omega_0=6.95886$ and $\omega_1=6.98631$. 
The former eigenmode is even and the latter one is odd.

We expect that the bifurcation shown in Fig. \ref{pcfig1} may occur, in the
nonlinear equation (\ref{pceq6}), at $\mu $ close to $-\omega_0$, and the
emerging asymmetric solution may be close to a superposition of the two
above-mentioned localized linear eigenmodes. We explore this
possibility by means of a Galerkin truncation based on these modes:
\begin{equation}
u(x,z)=c_{0}(z)v^{(0)}(x)+c_{1}(z)v^{(1)}(x),  \label{pceq8}
\end{equation}where $c_{0}$ and $c_{1}$ are assumed small. Substitution of (\ref{pceq8})
into (\ref{pceq6}), projecting onto the linear eigenmodes and retaining
leading-order nonlinear terms yields the following finite-dimensional
reduction:
\begin{eqnarray}
i\dot{c}_{0} &=&(\mu +\omega _{0})c_{0}-a_{00}|c_{0}|^{2}c_{0}-a_{01}\left(
2|c_{1}|^{2}c_{0}+c_{0}^{\star }c_{1}^{2}\right) ,  \label{pceq9} \\
i\dot{c}_{1} &=&(\mu +\omega _{1})c_{1}-a_{11}|c_{1}|^{2}c_{1}-a_{01}\left(
2|c_{0}|^{2}c_{1}+c_{1}^{\star }c_{0}^{2}\right) ,  \label{pceq10}
\end{eqnarray}where the overdot stands for $d/dz$, $a_{kl}~\equiv ~E_{0}\int_{-\infty
}^{+\infty }\left[ v^{(k)}(x)\right] ^{2}\left[ v^{(l)}(x)\right] ^{2}\left[
1+I_{0}(x)\right] ^{-2}dx$, and $a_{00}\approx 0.86302$, $a_{01}\approx
0.89647$, $a_{11}\approx 0.93958$.

Substituting further $c_{j}\equiv \rho _{j}\exp (i\phi _{j})$, and taking
into regard the conservation of the total norm, $\rho _{0}^{2}+\rho
_{1}^{2}=N$, we reduce Eqs. (\ref{pceq9}) and (\ref{pceq10}) to a system of
two real ordinary differential equations:
\begin{eqnarray}
\dot{\rho}_{0} &=&a_{01}\rho _{0}^{2}\rho _{1}\sin (2\Delta \phi ),
\label{pceq11} \\
\dot{\Delta \phi} &=&- \Delta \omega +a_{11}\rho _{1}^{2}-a_{00}\rho _{0}^{2}  \notag \\
&+&a_{01}\left( 2+\cos (2\Delta \phi )\right) \left( \rho _{0}^{2}-\rho
_{1}^{2}\right) ,  \label{pceq12}
\end{eqnarray}
where $\Delta \phi \equiv \phi _{1}-\phi _{0}$ and $\Delta \omega \equiv
\omega _{1}-\omega _{0}$. Since we are interested in real
solutions of the underlying equation (\ref{pceq6}), we
will confine our considerations to steady states with $\Delta \phi
=0$ ($\mathrm{mod}$ $\pi $). Then, from Eq. (\ref{pceq12}) one can
easily find that no solution bifurcates from $\rho _{0}=0$;
however, solutions with $\rho _{1}\neq 0$ can bifurcate from the
symmetric ones with $\rho _{0}^{2}=N$ and $\rho _{1}=0$.
These are the solutions that we are interested in, as they may
account for the SSB, due to the inclusion of the odd
eigenfunction $v^{(1)}(x)$ in Eq.
(\ref{pceq8}). The critical value of the norm at which the
bifurcation occurs, is found from Eqs. (\ref{pceq11}) and
(\ref{pceq12}) to be
\begin{equation}
N_{c}=\left( 3a_{01}-a_{00}\right) ^{-1}|\Delta \omega |.  \label{pceq13}
\end{equation}This simple prediction is the main finding of the analysis. One can also
find, from Eqs. (\ref{pceq9}) and (\ref{pceq10}), the critical propagation
constant $\mu _{c}$ at which the SSB is expected,
\begin{equation}
\mu _{c}=-\omega _{0}+a_{00}\left( 3a_{01}-a_{00}\right) ^{-1}|\Delta \omega
|.  \label{pceq14}
\end{equation}Equations (\ref{pceq13}) and (\ref{pceq14}) finally predict the occurrence
of the bifurcation at $N_{c}=0.01503$ and $\mu _{c}=-6.94589$, in remarkable
agreement with the numerical simulations reported above. In particular, the
relative error in $N_{c}$ is considerably less than $1\%$ in the worst case,
and the error in the prediction of $\mu _{c}$ is $\approx 0.002\%$.

%\section{Experimental Results}

\begin{figure}[t]
\centerline{
\includegraphics[width=\wwfig,height=\wwfig,angle=0,clip]{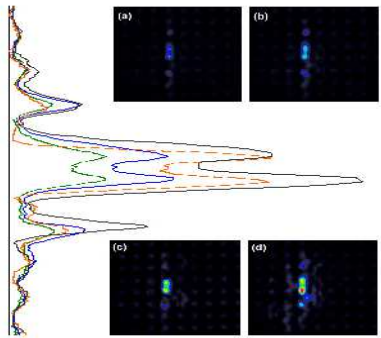}}
\caption{Experimental demonstration of SSB 
in an optically induced photonic lattice. From
(a) to (d), shown are the transverse intensity patterns of the
probe beam (initially a fundamental Gaussian beam) at an intensity
(normalized to the lattice intensity) of 0.1, 0.2, 0.3 and 0.5.
The bias field was kept at 2 kV/cm, and all other
parameters were fixed. The left side shows the 
corresponding vertical beam profiles.} \label{pclfig4}
\end{figure}

\textit{Experimental Results}. Experiments were performed
in a biased photorefractive crystal (SBN:60, 5x5x20 $mm^{3}$)
illuminated by a partially coherent beam periodically modulated
with an amplitude mask. This spatially modulated beam is
ordinarily polarized, so it experiences only a weak nonlinearity
and induces a stable square waveguide lattice
in the biased crystal \cite{MartinPRL04,ChenOL02}. The principal
axes of the lattice are oriented in horizontal and
vertical directions. A Gaussian beam, split off from the
same laser source, is used as a coherent probe beam. Contrary to
the experiments with discrete solitons
\cite{FleischerNature03,MartinPRL04,ChenPRL04}, the probe beam is
focused inter-site, i.e., between two lattice sites in
the vertical direction 
(to avoid any anisotropic effects 
such as those from self-bending, since the crystalline c-axis is oriented 
in the horizontal direction).
The probe beam is extraordinarily
polarized, propagating collinearly with the lattice through the
crystal. The polarization configuration is chosen so that
the probe would experience strong nonlinearity but the induced
lattice remains nearly undisturbed during propagation
\cite{EfremidisPRE02}. To demonstrate the SSB, all experimental
conditions were kept unchanged, except that the intensity of the
probe beam was increased gradually. Typical experimental results
are presented in Fig. \ref{pclfig4}. They were obtained
with a lattice of 45$\mu m$ nearest neighbor spacing. When the
intensity of the probe beam is low, its energy tunnels into two
waveguides symmetrically, 
as seen in the transverse patterns
of the probe beam 
in Fig. \ref{pclfig4}(a-c). Above a
threshold value of the input intensity, 
the intensity pattern of the probe beam at output becomes
asymmetric, as shown in Fig. \ref{pclfig4}(d). The bifurcation
from a symmetric to an  asymmetric output is also
clearly visible in the vertical profiles of the probe beam
displayed in Fig. \ref{pclfig4}. These results are
related to the even-mode solitons \cite{NeshevOL}, or the 
two parallel in-phase
solitons \cite{YangOL04}, observed previously in optically-induced
lattices, which were found to be unstable. Here we demonstrated a
clear bifurcation from symmetric to asymmetric states due to
the SSB, which is more relevant to the symmetry-breaking
instability of a two-humped, self-guided laser beam,
observed in a different nonlinear-optical system
\cite{HaeltermannPRL02}. We emphasize that the output patterns in
Fig. \ref{pclfig4} were all taken at steady state, and
the only parameter we varied was the intensity of the probe beam.

\textit{Conclusion and Discussion}. We have studied
numerically and analytically the
bifurcation from a symmetric to an asymmetric ground state in an
equation of the NLS type. The equation models the
propagation of a probe beam through an optically-induced
periodic lattice in a photorefractive nonlinear crystal. A
double-well potential locally approximates the spatial shape of
the lattice guiding the probe beam. The
ground-state profile breaks its symmetry at a
critical power, $N_{c}$. The transition is accompanied by
destabilization of the symmetric state. The threshold,
$N_{c}$, as well as other essential features of the SSB
bifurcation, are very well approximated by the
finite mode (Galerkin) approximation based on a
superposition of the symmetric and antisymmetric linear states of
the double-well potential. In parallel to the
theoretical analysis, we have reported the experimental observation of 
these phenomena in optically-induced photonic lattices in
a photorefractive crystal.

The analysis detailed herein can be adapted to models of
Gross-Pitaevskii or nonlinear-Hartree types, which are central to modeling
symmetry breaking and other phenomena in BECs \cite{frohlich,jackson,kutz}.
See also Ref. \cite{kapitula}, for such an analysis when a BEC is
loaded into a combination of a magnetic trap and an optical lattice.

Finally, although the simple two-mode truncation was used
here in the one-dimensional geometry, generalizations to
arbitrary dimensions and incorporation of higher-order
modes can be developed to describe 
more complex bifurcations, in both static and time-dependent settings,
as well as coupling to radiation waves \cite{mbs}. These
are directions currently under investigation.

This work was partially supported by NSF-DMS-0204585, NSF-CAREER,
and the Eppley Foundation for Research (PGK). AFOSR, ARO and NSFC
(ZC), the Israel Science Foundation through the
grant No. \ 8006/03 (BAM), and NSF-DMS-0412305 (MIW). We
are indebted to Todd Kapitula for valuable discussions.

\end{document}